\documentclass{jltp}
\usepackage{graphicx}

\newcommand{\ba}{\begin{eqnarray}}
\newcommand{\ea}{\end{eqnarray}}

\title{Total Energy Studies for Ferromagnetic Nickel: What is the Optimum Combination of the Multi-band Gutzwiller Method and Density Functional Theory?}

\author{T. Ohm, S. Weiser, R. Umst\"atter, and W. Weber\\
{\rm\em Institut f\"ur Physik, Universit\"at Dortmund,}\\
{\rm\em Otto-Hahn-Str.4, D-44221 Dortmund, Germany}\\
\and
and\\
J. B\"unemann \\
{\rm\em Physical and Theoretical Chemistry Laboratory, Oxford University,} \\
{\rm\em South Parks Road, Oxford OX1 3QZ, United Kingdom}}

\runninghead{T. Ohm, S. Weiser, R. Umst\"atter, W. Weber, and J. B\"unemann}{Total energy studies for ferromagnetic nickel}

\begin{document}

\maketitle

\begin{abstract}
The multi-band Gutzwiller method, combined with calculations based on density functional theory, is employed to study total energy curves of the ferromagnetic ground state of $Ni$. A new method is presented which allows flow of charge between $d$ and $s$, $p$ type orbitals in an approximate way. Further it is emphasized that the missing repulsive contribution to the total energy at large magnetic moments can be estimated from an analysis of specific DFT calculations.
\end{abstract}
\vspace{1em}


\section{INTRODUCTION}
Combining the multi-band Gutzwiller method\cite{prb98} and density functional theory (DFT) allows to study magnetic states of crystalline solids\cite{75birthd,wandlitz}. 
The presently used Gutzwiller DFT scheme employs a tight-binding model for the single particle energies derived from DFT calculations for non-magnetic crystal and uses certain chemical potentials which keep the partial charges $n_s$, $n_p$, and $n_d$ of $s$, $p$, and $d$ orbitals constant for all values of the magnetic moment $\mu$.
Calculations for ferromagnetic iron group metals, in particular for $Ni$, yield total energy versus magnetic moment curves $E_{tot}(\mu)$, which are in good agreement with experiment, yet the values $\mu_{min}$ at minimum total energy are somewhat too large.

In this paper we first present an alternate method which, in magnetic cases, allows charge flow between the various orbitals.
We also argue that additional repulsive contributions to the total energy exist near and beyond the limit of strong ferromagnetism, which are not included in our scheme.
These contributions can be estimated from an analysis of specific DFT calculations.
When incorporated in total energy curves, we obtain $\mu_{min}$ values very close to experiment.

\section{GROUND STATE ENERGY FOR NICKEL FROM THE MULTI-BAND GUTZWILLER METHOD }	
\label{section2}
\subsection{General treatment}

We start from the Hamiltonian
\ba
\label{starthamiltonian}
\hat{H} &=& \sum\limits_{\stackrel{i,j}{\sigma,\sigma^{\prime}}}\, t_{ij}^{\sigma\sigma^{\prime}}\, \hat{c}^\dagger_{i\sigma} \hat{c}_{j\sigma^{\prime}} + \sum\limits_i \, H_{at}^{(i)} \nonumber \\
&=& \hat{H}_1 + \hat{H}_{at}
\ea
Here, the quantities $t_{ij}^{\sigma\sigma^{\prime}}$ are the matrix elements of the tight-binding Hamiltonian $\hat{H}_1$.
For $Ni$ we use a basis of $4s$, $4p$, and $3d$ orbitals, in total $2 \times 9 = 18$ spin-orbitals $\sigma$.
The $t_{ij}^{\sigma\sigma^{\prime}}$ values have been obtained from a least squares fit to the energy bands of a DFT calculation with the LAPW-WIEN code for non-magnetic $Ni$, using the local density approximation (LDA)\cite{wien}.
In the fits we have preferably used energies of $k$ points at high symmetry points and along high symmetry lines.
This was done in such a way that the full information on the symmetry of all states involved in the fit was taken into account\cite{matweb80}.
The procedure allows to incorporate information on the orbital character of the states.

The atomic Hamiltonians
\ba
H_{at}^{(i)} &=& \sum\limits_{\sigma_1 ... \sigma_4}\, U^{\sigma_1 ... \sigma_4}\, \hat{c}^\dagger_{\sigma_1}\hat{c}^\dagger_{\sigma_2}\hat{c}_{\sigma_3}\hat{c}_{\sigma_4} \nonumber \\
&=& \sum\limits_\Gamma \, E_\Gamma \, \left| \Gamma\right>\left<\Gamma  \right|
\ea
are assumed to be the same for each lattice site $i$ and include all on-site Coulomb interactions within the $3d$ shell.
We use the spherical atom approximation so that all interactions are determined by the three Slater-Condon integrals $F^0$, $F^1$, $F^2$ or, equivalently, by the three Racah parameters $A$, $B$, $C$ as outlined, e.g. in the textbook of Sugano {\em et al} \cite{sugano}.

In the subspace of the $3d$ shell, the states $\left| \Gamma\right>$ represent all $2^{2\cdot 5} - (2\cdot 5 + 1)$ multi-electron eigenstates with eigenenergies $E_\Gamma$.\\
The expectation value of $\hat{H}$ is given by\cite{prb98}
\ba
\label{expecH}
\left<\hat{H}\right> &=& \sum\limits_{\stackrel{i,j}{\sigma,\sigma^{\prime}}}\, t_{ij}^{\sigma\sigma^{\prime}}\, \sqrt{q_\sigma}\, \sqrt{q_{\sigma^{\prime}}}\, \left<\Phi_0 \left| \hat{c}^\dagger_{i \sigma} \hat{c}_{j \sigma^{\prime}} \right|\Phi_0\right> \nonumber \\
&+& \sum\limits_{i \sigma}\, (1 - q_\sigma)\, \epsilon_\sigma\, n_\sigma + \sum\limits_{i \Gamma}\, E_\Gamma\, m_\Gamma \nonumber \\
&=& \left<\Phi_0 \left| \hat{H}_{eff}\right| \Phi_0\right>.
\ea
Here, 
\vspace{-0.3em}
\ba
 n_\sigma =n_{i\sigma} = \left< \Phi_0\left|\hat{n}_{i\sigma}\right|\Phi_0\right> 
\ea
is the density in the spin-orbital $\sigma$ (we assume a monoatomic cubic lattice and skip all labels $i$, $j$, if possible) and $\epsilon_{\sigma} =\epsilon_{i\sigma} =  t_{ii}^{\sigma\sigma}$. The expectation value (\ref{expecH}) has to be minimized with respect to the single particle product wave function $\left|\Phi_0\right>$ and the occupancies $m_\Gamma$ of the atomic multi-electron states $\left|\Gamma\right>$. Both determine the 'hopping reduction factors' $q_\sigma$ via the relations 
\ba
\sqrt{q_\sigma} = \sum\limits_{\Gamma \Gamma^{\prime}}\, S^\sigma_{\Gamma \Gamma^{\prime}}\, \sqrt{m_\Gamma}\, \sqrt{m_{\Gamma^{\prime}}}
\ea
where the coefficients $S^\sigma_{\Gamma \Gamma^{\prime}}$ depend on the densities $n_{\sigma^\prime}$ and the eigenstates \mbox{$\left|\Gamma\right>$} of $H_{at}^{(i)}$.
 $\left|\Phi_0\right>$ is the ground state of $\hat{H}_{eff}$, but it has to be determined in a self-consistent way, since $\hat{H}_{eff}$ depends on  $\left|\Phi_0\right>$  via the densities $n_\sigma$.

For the optimum set of variational parameters, the effective single particle Hamiltonian leads to quasi-particle energy bands of a Fermi liquid which determine, e.g., the shape of the Fermi surface or may be compared to experimental energy bands, obtained from angular resolved photoemission.

The wave function $\left|\Phi_0\right>$ can be chosen to yield magnetic ground states by incorporating orbital exchange splittings $\Delta t_{2g}$ and $\Delta e_g$ to produce majority and minority bands.
Then $\left| \Phi_0^{opt}\right>$ is a function of the magnetic moment $\mu$.
In fact, for a given value of the magnetic moment $E_{tot}(\Delta t_{2g},\Delta e_g)$ shows a minimum, when $\Delta t_{2g} / \Delta e_g \approx 2$-$3$; i.e. the exchange splitting is strongly anisotropic.

\subsection{The method of chemical potentials}

\begin{figure}
\centerline{\includegraphics[width=.7\textwidth,clip=true]{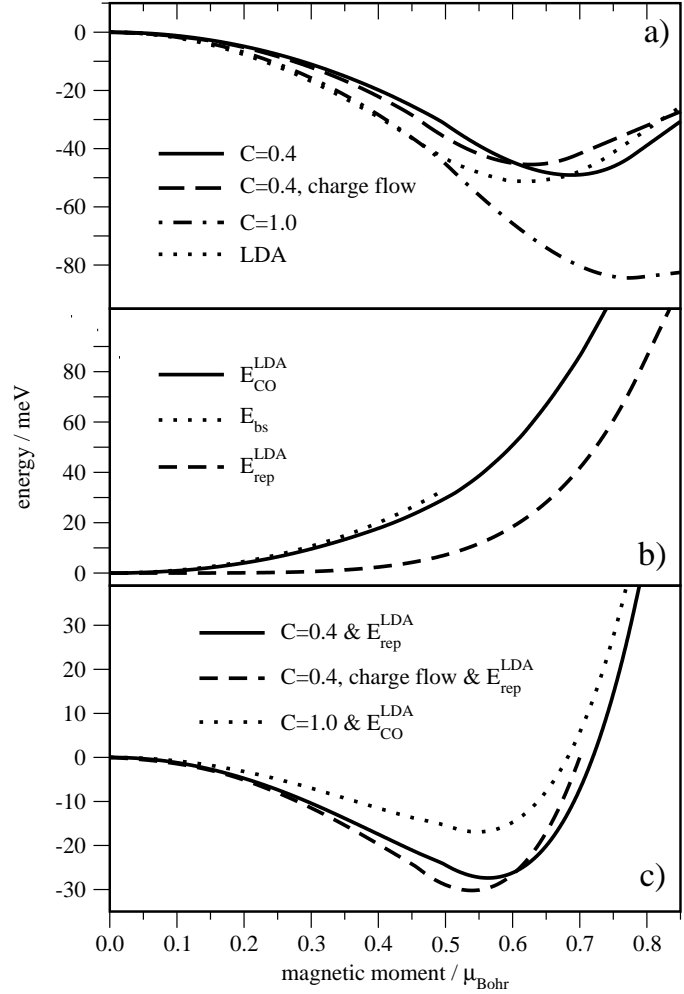}}
\caption{Gutzwiller DFT total energy curves versus magnetic moment $\mu$ for ferromagnetic $Ni$. 
$a)$ shows calculations using the chemical potential scheme (exchange interactions $C = 0.4\, \mathrm{eV}$ and $C = 1.0\, \mathrm{eV}$, respectively) and using the new charge flow method ($C = 0.4\, \mathrm{eV}$). 
Also shown is the LDA total energy curve. 
$b)$ exhibits the repulsive part of the LDA total energy curve $E_{CO}^{LDA}$, obtained by keeping only the charge-dependent part of the exchange correlation potential. 
Further shown is the one-particle band-splitting contribution $E_{bs}$ at small moments $\mu$ and the residual repulsive energy curve $E_{rep}^{LDA}$. 
c) shows the three Gutzwiller DFT total energy curves of a), corrected by either $E_{rep}^{LDA}$ (for the $C=0.4\, \mathrm{eV}$ calculations) or by $E_{CO}^{LDA}$ (for the $C=1.0\, \mathrm{eV}$ calculation).}
\label{figure1}
\end{figure}

The unlimited variation of the spin-orbital charge densities $n_\sigma$ leads to a charge flow from $3d$ to $4s$ and $4p$ states, since the Hamiltonian of eq. (\ref{starthamiltonian}) includes only atomic $3d$ electron-electron interactions.
The charge flow can be suppressed by appropriate chemical potentials which keep the values of the charge densities fixed to specific values $n_s^{DFT}$, $n_p^{DFT}$, and thus also $n_d^{DFT}$.
These values are obtained from the ground state of $\hat{H}_1$, the tight-binding model, which is derived from the DFT results for non-magnetic $Ni$.
The choice is justified by the observation that, in general, charge densities obtained from DFT calculations show very good, albeit not perfect, agreement with experimental data.
Typical total energy curves $E_{tot}(\mu)$ for $Ni$ are shown in Fig. \ref{figure1}a.
The first calculation has been carried out using $A=10\, \mathrm{eV}$, $C=0.4\, \mathrm{eV}$ and $C/B=4.5$.
$\mu_{min} \approx 0.68$ is found, and $E_{cond} = E_{tot}(\mu$=$0) - E_{tot}(\mu_{min}) \approx 40\, \mathrm{meV}$.
Note that the ratio $C / B$ is typical for $3d$ ions and that the exchange interaction parameter $C = 0.4\, \mathrm{eV}$ is an estimate derived from free $Ni$ ions\cite{sugano}.
The Racah parameter $A$ corresponds to the Hubbard model parameter $U$; its value is adjusted to produce the experimental $d$-band width of $Ni$ \cite{75birthd,wandlitz}.

The second calculation employs a significantly larger $C=1.0\, \mathrm{eV}$, again with $C/B = 4.5$ and $A = 10\, \mathrm{eV}$.
Here we obtain $\mu_{min} = 0.77$ and $E_{cond} \approx 84\, \mathrm{meV}$.

Our value for the Racah parameter $A=10\, \mathrm{eV}$ is much larger than suggested in other works (see, e.g., Refs. \onlinecite{pruschke,kotliar01}). 
This discrepancy may have to do with the neglect of hybridization terms between $s$, $p$ and $d$ orbitals\cite{pruschke}. 
If we switch off these terms in our theory we find $A\approx 3\, \mathrm{eV}$ which is comparable to the results in Ref. \onlinecite{kotliar01}.

\subsection{Approximate incorporation of other contributions to the electron-electron interaction: the charge flow method} 

The charge flow can be allowed if additional interaction terms involving the $s$ and $p$ type charge densities are included in the total energy.
A simple term is of the form
\ba
E_{sp} = A_{sp,d}\, (n_s + n_p)\, n_d + \frac{1}{2}\, A_{sp}\, (n_s + n_p)^2
\ea

If we add $E_{sp}$ to the Gutzwiller ground state energy for paramagnetic nickel, now unrestricted in the charge densities, we can choose $A_{sp,d}$ and $A_{sp}$ in such a way as to yield the minimum total energy for the densities $n_s^{DFT}$ and $n_p^{DFT}$.
Typical values of $A_{sp,d}$ and $A_{sp}$ are of order of the Racah parameter $A$.
Introducing magnetic states $\left| \Phi_0 (\Delta t_{2g},\Delta e_g)\right>$, the changes in $s$-$d$ and $p$-$d$ hybridization due to the exchange splittings $\Delta t_{2g}$, $\Delta e_g$ of majority and minority bands leads to increasing $n_s$, $n_p$ values (and decreasing $n_d$).
These changes cause $E_{sp}$ to increase and compensate for the reduction of the $d$-$d$ interaction energy (which in first approximation is $\frac{1}{2}\, A\, (n_d)^2$).
A typical total energy curve (using $A = 10\, \mathrm{eV}$, $C = 0.4\, \mathrm{eV}$ and $C/B = 4.5$) is also shown in fig. \ref{figure1}a.
The curve is quite similar to the corresponding $E_{tot}(\mu)$ curve using the chemical potential method, yet the value $\mu_{min} \approx 0.625\, \mu_B$ is somewhat smaller than before.

\section{ANALYSIS OF DFT TOTAL ENERGY CALCULATIONS} 

Using spin-dependent DFT, again within the LDA, we reproduce the well known $E_{tot}^{LDA}(\mu)$ curve with its minimum at $\mu_{min}^{LDA} = 0.62\, \mu_B$ (also see fig. \ref{figure1}a).
If we put equal to zero the spin-dependent part of the LDA exchange correlation functional, we obtain a repulsive curve $E^{LDA}_{XC,\, Charge\, Only}(\mu)$ (fig. \ref{figure1}b).
This curve $E_{CO}^{LDA}(\mu)$ has its minimum at $\mu =0$ and coincides there with $E_{tot}^{LDA}(\mu =0)$, as expected.
The function $E_{CO}^{LDA}(\mu )$ can be interpolated very well by the polynomial
\ba
E_{CO}^{LDA}(\mu) - E_{tot}^{LDA}(0) = \alpha_2 \mu^2 + \alpha_4 \mu^4 + \alpha_6 \mu^6,
\ea
with $\alpha_2 = 0.090\ \mathrm{eV}/\mu_B^2$, $\alpha_4 = 0.054\ \mathrm{eV}/\mu_B^4$, $\alpha_6 = 0.244\ \mathrm{eV}/\mu_B^6$.

The term $\alpha_2 \mu^2$ dominates $E_{CO}^{LDA}$ in the range values $0 \le \mu < 0.4$; i.e. close to the limit of strong ferromagnetism, where the majority $d$ bands have been filled completely.
The term $\alpha_2 \mu^2$ appears to arise completely from the splitting of the majority and minority bands.
This can be shown by evaluating the sum over the LDA valence band energies.
For the non-magnetic case, the tight-binding model of $\hat{H}_1$ agrees very well with the LDA energy bands i.e., $E_{bs} = \left<\Phi_0^{(1)}\left| \hat{H}_1\right| \Phi_0^{(1)}\right>$, with $\left|\Phi_0^{(1)}\right>$ being the ground state wave function of $\hat{H}_1$, gives the sum of all occupied LDA valence band energies. 
For small values of the exchange splittings $\Delta t_{2g}$, $\Delta e_g$ (and thus small values of the magnetic moment $\mu$) the majority and minority LDA $d$-bands shift rigidly, which is very well reproduced by corresponding $\left| \Phi_0^{(1)}(\Delta t_{2g},\ \Delta e_g)\right>$ states.
Thus, for small $\mu$, the expression
\ba
E_{bs}(\mu) &=& \left<\Phi_0^{(1)} (\Delta t_{2g}\mbox{, } \Delta e_g) \left| \hat{H}_1 \right| \Phi_0^{(1)}(\Delta t_{2g}\mbox{, } \Delta e_g)\right> \nonumber \\
&=& \alpha_0^{(1)} + \alpha_2^{(1)} \mu^2 + ... 
\ea
should be compared to $E_{tot}^{LDA,\, CO}(\mu)$.
Indeed, we find $\alpha_2^{(1)} \approx \alpha_2$.
However, the term $E_{bs}(\mu)$ i.e. the increase in total energy due to the splitting of minority and majority bands is included in the Gutzwiller treatment eq.(\ref{expecH}). 
Therefore, the repulsive term
\ba
E_{rep}^{LDA} (\mu) = \alpha_4 \mu^4 + \alpha_6 \mu^6
\ea
\begin{figure}[p]
\centerline{\includegraphics[height=\textwidth,angle=270]{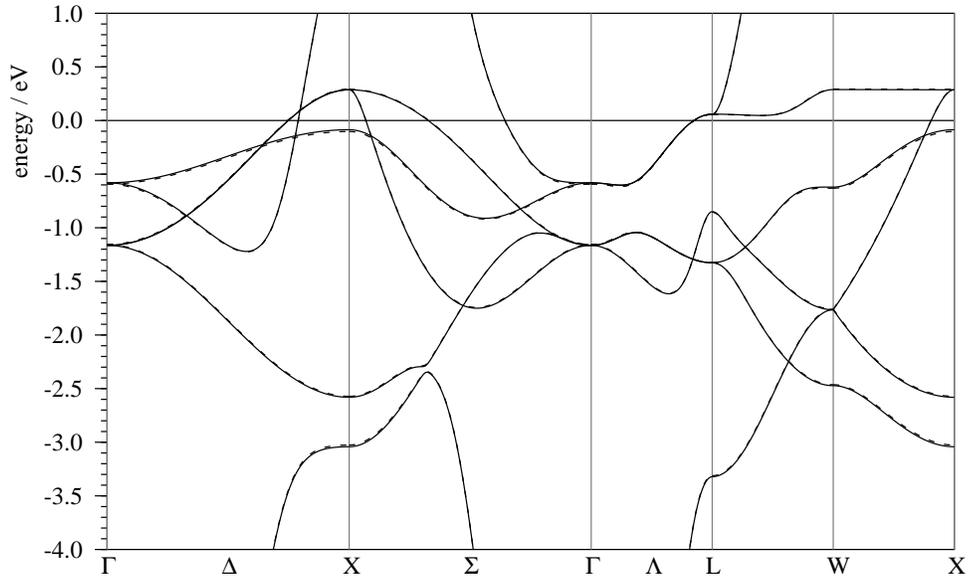}}
\caption{Minority quasi particle bands for ferromagnetic $Ni$ ($\mu = 0.55\ \mu_B$), obtained from the Gutzwiller DFT calculations using either $C = 0.4\, \mathrm{eV}$ (full lines) or $C = 1.0\, \mathrm{eV}$ (dashed lines). The two calculations differ only on a scale of a few $\mathrm{meV}$.}
\label{figure2}
\end{figure} 
\begin{figure}[p]
\centerline{\includegraphics[height=\textwidth,angle=270]{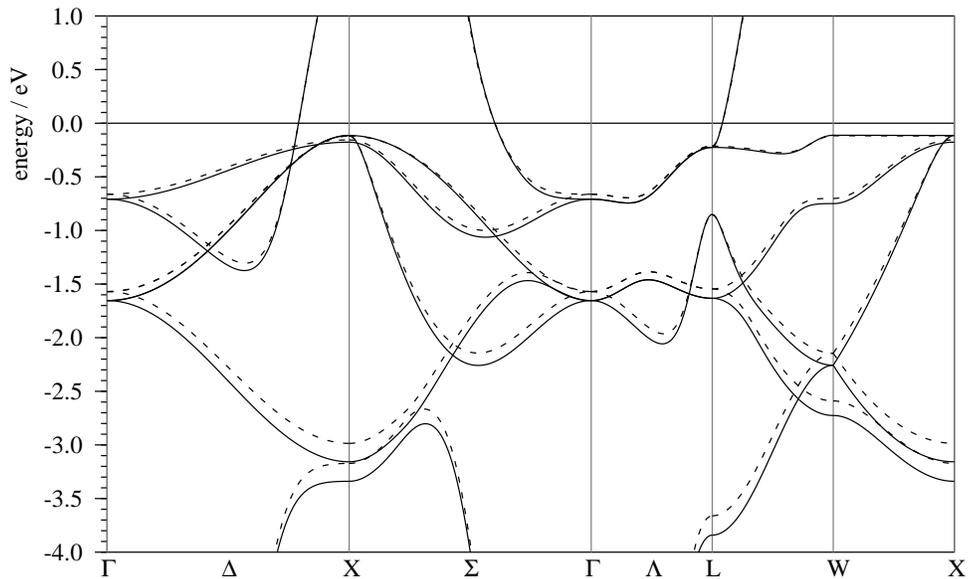}}
\caption{Majority quasi particle bands for ferromagnetic $Ni$ ($\mu = 0.55\ \mu_B$) from the $C = 0.4\, \mathrm{eV}$ (full lines) and the $C = 1.0\, \mathrm{eV}$ calculations (dashed lines). The two calculations differ on a scale of $10$-$100\, \mathrm{meV}$.}
\label{figure3}
\end{figure} 
is interpreted to represent the missing repulsive part of our Gutzwiller treatment.
We remind our readers that the Hamiltonian of eq. (\ref{starthamiltonian}) is still a very simplified model. It uses a very limited local basis, assumes the orbitals of the neighboring atoms to be orthogonal, only valence electrons are considered and only part of the electron-electron interaction is included.
On the other hand, density functional theory treatments have a rather complete basis, incorporate all core states and all electron-electron interactions, albeit on the basis of an effective single particle theory.

If we add $E_{rep}^{LDA}$ to the total energy curves, obtained either using the chemical potential method or the charge flow method, we find very good agreement of $\mu_{min}$ with the experimental spin-only value of $0.55\, \mu_B$ (see fig. \ref{figure1}c).

Alternatively, if we interprete the full curve $E_{CO}^{LDA}(\mu)$ as the repulsive part missing in the Gutzwiller treatment, the curves using $C = 0.4\, \mathrm{eV}$ do not give any minimum for magnetic states.
Only when we enhance $C$ to $C = 1\, \mathrm{eV}$, a reasonable value of $E_{cond}$ is found, again for $\mu_{min}$ close to the experimental value.

\section{QUASI-PARTICLE ENERGY BANDS}

The change of $C$ from $0.4\, \mathrm{eV}$ to $1.0\, \mathrm{eV}$ may not only produce quite different total energy curves, but might also result in drastic changes of the quasi particle energy bands.
Therefore in fig. \ref{figure2} and fig. \ref{figure3} the corresponding bands for $C = 0.4\, \mathrm{eV}$, and $C = 1.0\, \mathrm{eV}$ are compared, both calculated at $\mu = 0.55\, \mu_B$ and after minimizing the total energy with respect to the ratio $\Delta t_{2g} / \Delta e_g$ (which is rather similar for the two cases).
We find few differences of the quasi particle bands, very little differences concerning the minority bands (see fig. \ref{figure2}) and somewhat bigger ones for the majority bands (see fig. \ref{figure3}).
This result is not surprising, since even for $C = 1.0\, \mathrm{eV}$, we are still in the limit $C / A \ll 1$, where the $q_\sigma$-values which enter the effective single particle Hamiltonian are mainly determined by $A$.
The quasi-particle bands are also influenced by the exchange splittings $\Delta t_{2g}$ and $\Delta e_g$.
Again, since we find similar ratios $\Delta t_{2g} / \Delta e_g$ for the two cases, the anisotropy of the exchange splitting is similar.

\section{CONCLUDING REMARKS}

We have presented an alternate scheme for carrying out Gutzwiller calculations which allows charge flow.
This scheme uses simple approximations to treat in the total energy expression those electron-electron interaction terms, which are not included in the Gutzwiller Hamiltonian. The interaction parameters are determined from paramagnetic calculations. This new scheme appears to be somewhat superior to the previously introduced chemical potential method, where all charges are kept constant.
In both methods, the magnetic moment values at minimum total energy are somewhat larger than the corresponding experimental spin-only moment.

We have tried to estimate the missing repulsive contribution to the total energy from an analysis of specific DFT calculations, where the spin-dependent part of the exchange-correlation potential has been switched off.
When this repulsive term is corrected for the band splitting contributions already included in the Gutzwiller method, the resulting total energy curves yield values of the magnetic moments which agree very well with experiment.

If we assume that the repulsive part of the LDA total energy curves remains uncorrected, i.e. if the band-splitting contribution is kept in the repulsive term, calculations using a bigger interaction parameter $C$=$1\, \mathrm{eV}$ lead to very satisfactory results concerning the magnetic moment.
We then compare the quasi particle energy bands based on the two calculations with $C$ values of $0.4\, \mathrm{eV}$ and $1.0\, \mathrm{eV}$, respectively. 
Both calculations yield the same (experimental) value of the magnetic moment. 
We find very little differences in the energies of minority bands and only moderate differences for the majority bands.

These results justify a simplified Gutzwiller DFT treatment, where the magnetic moment is fixed to the experimental spin-only value and $E_{tot}$ is minimized with respect to the ratio $\Delta t_{2g} / \Delta e_g$ of the orbital exchange fields.

%
%


\begin{thebibliography}{9}

\bibitem{prb98}
J.~B\"unemann, W.~Weber, and F.~Gebhard
Phys.~Rev.~B {\bf 57}, 6896 (1998).

\bibitem{75birthd}
J.~B\"unemann, F.~Gebhard, and W.~Weber, 
Found. of Physics {\bf 30},  2011 (2000)

\bibitem{wandlitz}
W.~Weber, J.~B\"unemann, and F.~Gebhard, 
in K.~Baberschke, M.~Donath, W.~Nolting, eds., 
{\em Band-Ferromagnetism}, Lecture Notes in Physics (Springer, Berlin, in press).

\bibitem{wien}
P.~Blaha, K.~Schwarz, P.I.~Sorantin, and S.B.~Trickey, Comp. Phys. Commun. {\bf 59}, 399 (1990).

\bibitem{matweb80}
W.~Weber and L.F.~Mattheiss,
Phys. Rev. B {\bf 25}, 2270 (1982).

\bibitem{sugano}
S.~Sugano, Y.~Tanabe, and H.~Kamimura, {\em Multiplets of Transition-Metal Ions in Crystals}, Pure and Applied Physics {\bf 33} (Academic Press, New York, 1970)

\bibitem{pruschke}
T.~Pruschke,
private communication.

\bibitem{kotliar01}
A.I.~Lichtenstein, M.I.~Katsnelson, and G.~Kotliar,
Phys. Rev. Lett. {\bf 87}, (2001) in press.

\end{thebibliography}
\end{document}